\documentclass[conference]{IEEEtran}
\IEEEoverridecommandlockouts
\usepackage{cite}
\usepackage{amsmath,amssymb,amsfonts}
\usepackage{algorithm}
\usepackage{booktabs}
\usepackage{algpseudocode}
\usepackage{graphicx}
\usepackage{textcomp}
\usepackage{xcolor}
\usepackage{multirow}
\usepackage{multicol}
\usepackage{url}
\usepackage[a4paper, total={184mm,239mm}]{geometry}
\def\BibTeX{{\rm B\kern-.05em{\sc i\kern-.025em b}\kern-.08em
    T\kern-.1667em\lower.7ex\hbox{E}\kern-.125emX}}
\begin{document}

\title{Self-Adaptive Ising Machines for Constrained Optimization

}

\author{\IEEEauthorblockN{ Corentin Delacour}
\IEEEauthorblockA{\textit{Department of Electrical and Computer Engineering} \\
\textit{University of California Santa Barbara}\\
delacour@ucsb.edu}
}

\maketitle

\begin{abstract}
Ising machines (IM) are physics-inspired alternatives to von Neumann architectures for solving hard optimization tasks. By mapping binary variables to coupled Ising spins, IMs can naturally solve unconstrained combinatorial optimization problems such as finding maximum cuts in graphs. However, despite their importance in practical applications, \textit{constrained} problems remain challenging to solve for IMs that require large quadratic energy penalties to ensure the correspondence between energy ground states and constrained optimal solutions. To relax this requirement, we propose a \textit{self-adaptive} IM that iteratively shapes its energy landscape using a Lagrange relaxation of constraints and avoids prior tuning of penalties. Using a probabilistic-bit (p-bit) IM emulated in software, we benchmark our algorithm with multidimensional knapsack problems (MKP) and quadratic knapsack problems (QKP), the latter being an Ising problem with linear constraints. For QKP with 300 variables, the proposed algorithm finds better solutions than state-of-the-art IMs such as Fujitsu's Digital Annealer and requires 7,500x fewer samples. Our results show that adapting the energy landscape during the search can speed up IMs for constrained optimization.
\end{abstract}

\begin{IEEEkeywords}
Ising Machines, Constrained Optimization, Lagrange relaxation, Knapsack problems, probabilistic bit
\end{IEEEkeywords}

\section{Introduction}
\begin{figure}[t!]
    \centering
    \includegraphics[width=\linewidth]{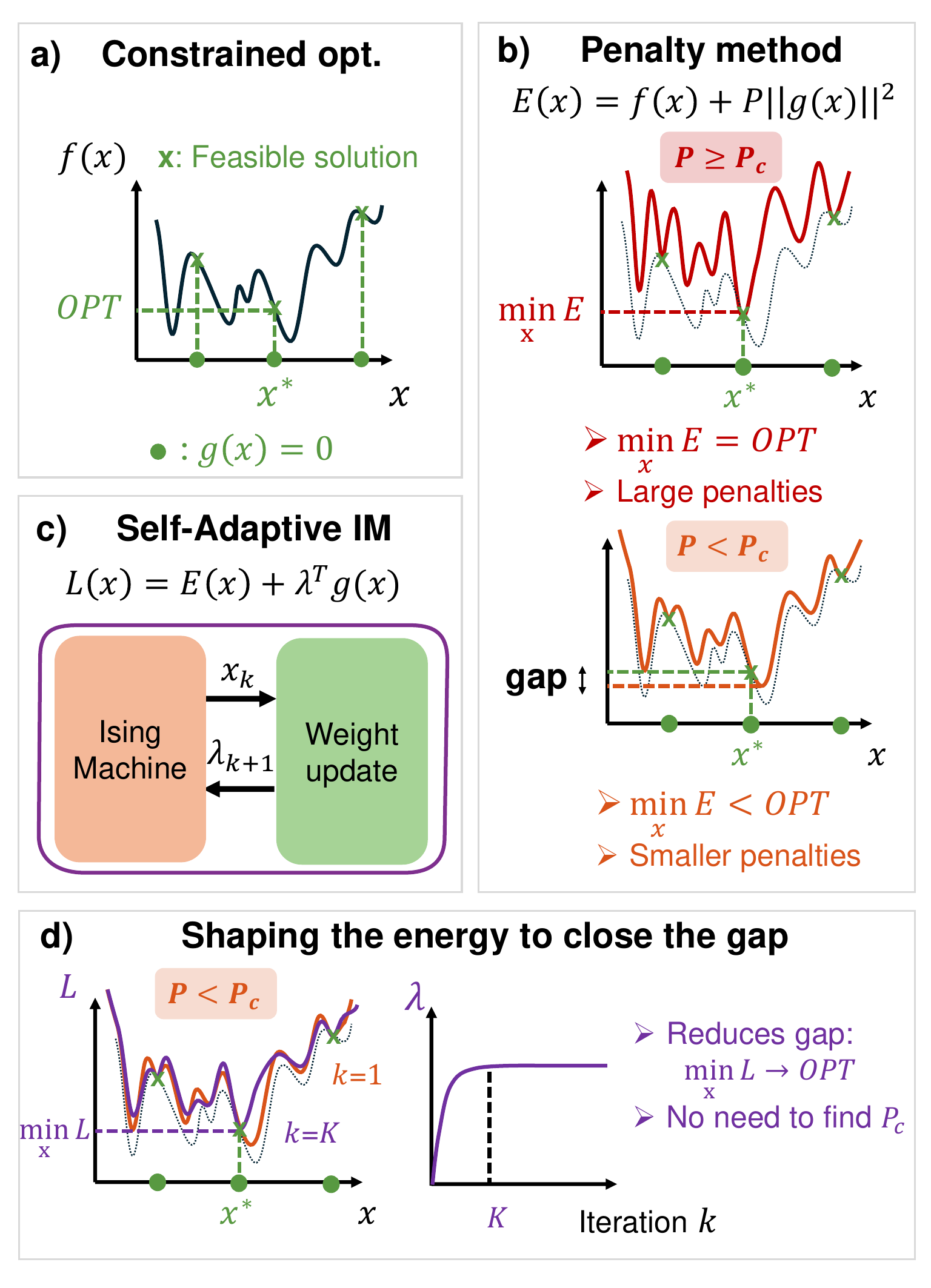}
    \caption{a) A constrained optimization problem consists of finding a minimum of $f(x)$ among feasible states defined by $g(x)=0$ (green dots). b) QUBO mapping for Ising machines. The classical penalty method induces positive penalties in the energy landscape. $P$ determines the penalty magnitude and needs to be larger than a critical value $P_C$ to ensure $\min_xE=OPT$. A smaller $P$ favors $f(x)$ but produces a gap $OPT-\min_xE>0$ where $\min_xE$ is unfeasible. c) To close the gap while keeping reasonable energy penalties, we propose a self-adaptive IM that autonomously adjusts its energy landscape based on a Lagrange relaxation of constraints. d) From an arbitrary $P<P_C$, the algorithm iteratively shapes the energy landscape using measured samples and is compatible with any programmable IM.}
    \label{self_adaptive_IM}
\end{figure}
Owing to their parallel computing capability, Ising machines (IMs) hold great promises for accelerating challenging optimization problems \cite{mohseni_2022}. Various technologies based on optics \cite{mcmahon_2016,honjo_2021}, memristors \cite{fahimi_2021,jiang_2023}, coupled oscillators \cite{wang_2021,Dutta_2021,bashar_2021}, digital electronics \cite{aramon_2019,aadit_2022} or quantum annealers \cite{king_2023} are currently being explored to implement IMs in hardware. The key principle is to harness the dynamics of distributed processing units $S_i=\pm1$ called \textit{spins}, that \textit{naturally} minimize the Ising Hamiltonian:
\begin{equation}
    H=-\sum_{i<j}J_{ij}S_iS_j-\sum_ih_iS_i
    \label{Ising_hamiltonian}
\end{equation}
where $J_{ij}$ and $h_i$ are the coupling elements and spin fields, respectively. Since the Ising decision problem (does the ground state of $H$ is negative?) is nondeterministic polynomial-time (NP)-complete \cite{lucas_2014}, the hope is that some hard combinatorial optimization problems could be solved more efficiently using IMs than classical von Neumann architectures. For instance, minimizing (\ref{Ising_hamiltonian}) is equivalent to the NP-hard problem of maximizing the cut of a graph where its vertices correspond to spins and graph edges are weighted by $W_{ij}=-J_{ij}$ \cite{lucas_2014}.

IMs have already shown promising results in solving quadratic unconstrained binary optimization problems (QUBO) such as max-cut \cite{honjo_2021,wang_2021}. However, solving \textit{constrained} problems with IMs has been less studied in the literature despite the need to model constraints in various real-life applications \cite{wolsey_2020}. To name a few, constraints on limited resources are found in capital budgeting, portfolio optimization, or production planning \cite{wilbaut_2007}. Constraints can also model forbidden paths in vehicle routing problems or impose sequences of operations for job-shop scheduling problems \cite{wolsey_2020}. Unfortunately, IMs do not naturally support constraints and are principally designed to minimize a cost function mapped to (\ref{Ising_hamiltonian}). The state-of-the-art approach to handle constraints with IMs consists of positively penalizing the Ising energy (\ref{Ising_hamiltonian}) when a state $S$ is not feasible (unsatisfied constraints) and is called \textit{the penalty method} \cite{lucas_2014,jimbo_2022,Bontekoe_2023}.
Despite the simplicity of the method, finding the optimal amount of energy penalty or deriving tight bounds is a hard problem that often leads to a tuning phase and worsens the global execution time \cite{lucas_2014,parizy_2021,Bontekoe_2023}.

In this paper, we propose to relax this requirement by harnessing a Lagrange relaxation of constraints \cite{fisher_2004} which allows the IM to automatically find the optimal energy penalties in a \textit{self-adaptive} manner. The key idea is to start the self-adaptive IM (SAIM) with some initial energy penalties, and iteratively refine them after each measurement so that the ground state of (\ref{Ising_hamiltonian}) tends to a constrained optimal solution.

After describing the classical penalty method, we introduce the SAIM algorithm and benchmark it with two hard-constrained optimization problems: the quadratic knapsack problem (QKP) and the multidimensional knapsack problem (MKP) using a probabilistic-bit (p-bit) IM emulation in software. Our results show that adapting the energy landscape during the IM operation increases the accuracy and reduces the number of samples by at least two orders of magnitude compared to state-of-the-art IMs.

\section{Constrained Optimization with Ising Machines}
We focus on constrained optimization problems whose optimal values are expressed as:
\begin{align}
   OPT= &\min_x f(x) \label{primal_problem} \\ \nonumber
    & \text{s.t. } g(x)=0 \nonumber
\end{align}
where $f: x\in\{0;1\}^N\rightarrow \mathbb{R}$ is an objective function to minimize and $g: x\in\{0;1\}^N\rightarrow \mathbb{R}^M$ is a constraint function satisfied when $g(x)=0$. For a standard IM with quadratic interactions, $f$ is at most quadratic and $g$ is linear. However, one could design a high-order IM supporting higher polynomial degrees for $f$ and $g$ \cite{bybee_2023}.
\subsection{Penalty Method}
 The penalty method adds the constraints weighted by a real parameter $P>0$ to the objective function as:
\begin{equation}
    E=f(x)+P\,||g(x)||^2
    \label{penalty_energy}
\end{equation}
The second term \textit{penalizes} the energy $E$ when $x$ lies in an unfeasible region ($g(x)\neq0$) and is illustrated in Fig.\ref{self_adaptive_IM}b. Intuitively, a large $ P$ value favors feasible states and makes $f(x)$ negligible in $E$, challenging the search for $OPT$. Conversely, a small $P$ value weakens the constraints and favors unfeasible $x$ at low energy. The minimization process corresponds to:
\begin{equation}
   LB_P= \min_x E
   \label{min_E}
\end{equation}
where $LB_P$ is a lower bound on the original problem (\ref{primal_problem}) since we have $LB_P=\min_xE\leq E(x^*)=OPT$ with $x^*$ an optimal solution for (\ref{primal_problem}). The equality $LB_P=OPT$ occurs for sufficiently large $P\geq P_C$ where $P_C$ is a critical value that depends on each instance \cite{lucas_2014}. Thus, there is a clear trade-off between accuracy and feasibility since with small $P$ values the IM is more likely to find lower bounds that are unfeasible by definition, and with large $P$ the rugged energy surface induces hard-to-escape local minima \cite{parizy_2021,jimbo_2022}.  
However, we show next that it is possible to relax the $P\geq P_C$ requirement by adding a linear contribution of constraints.

\subsection{Lagrange Relaxation for the Penalty Method}
\begin{figure}[t]
    \centering
    \includegraphics[width=\linewidth]{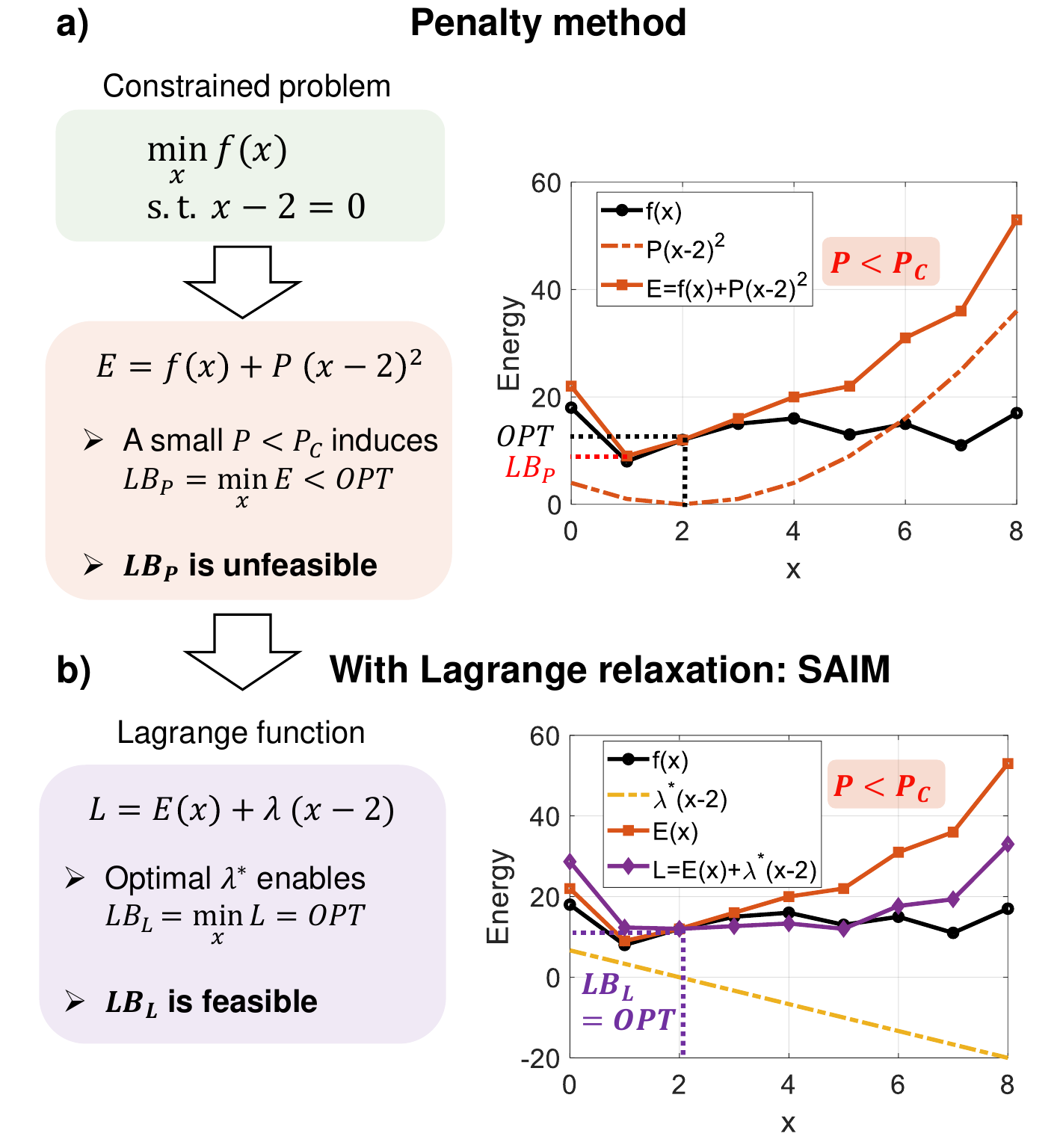}
    \caption{Illustration of the penalty method and the Lagrange relaxation for a minimization problem subjected to a toy constraint $x=2$ where $P$ is smaller than the critical value $P_C$. a) A small $P<P_C$ is insufficient for the penalty method to satisfy $LB_P=OPT$ and the minimization process is likely to provide the unfeasible $LB_P$. b) Adding a Lagrange relaxation of constraints allows $L$ to close the gap for $\lambda^*$ where $LB_L=OPT$.}
    \label{LR_illustration}
\end{figure}
To avoid finding $P\geq P_C$ that can lead to impractical energy penalties, we propose to harness a Lagrange relaxation of constraints $g(x)$ with some $P<P_C$. Compared to the penalty method, the Lagrange relaxation \textit{linearly} weights constraints by \textit{Lagrange multipliers} $\lambda\in \mathbb{R}^m$ and adds them to the energy function as \cite{fisher_2004}:
\begin{equation}
    L=E(x)+\lambda^T\,g(x)
    \label{Lagrange_energy}
\end{equation}
For readers more familiar with IMs, the purpose of coefficients $\lambda$ can be introduced in this way. Consider that by some mechanism, $\lambda$ are variables evolving in time at a slower rate than the minimization process defined as:
\begin{equation}
   LB_L= \min_x L
    \label{min_L} 
\end{equation}
which provides $\Bar{x}=\text{argmin}_x L$. If $g(\Bar{x})\neq 0$, $\Bar{x}$ is unfeasible, and one way of penalizing the energy $L$ would be to decrease (increase) $\lambda$ if $g(\Bar{x})$ is negative (positive) where variables $\lambda$ act as adjusting forces to move the minimum $\Bar{x}$ into a feasible region. Possible $\lambda$-dynamics could be:
\begin{equation}
    \tau_\lambda \frac{d\lambda}{dt}=g(\Bar{x})=+\nabla_\lambda LB_L
\end{equation}
where $\nabla_\lambda$ is a generalization of gradient defined for nondifferentiable functions \cite{xing_zhao_1997,fisher_2004}. Note how this process can be interpreted as the maximization of the lower bound $LB_L$ with respect to $\lambda$. For an arbitrary $\lambda$, $LB_L$ can be smaller than $OPT$ and the goal is to find the optimal $\lambda^*$ such that ideally $LB_L=OPT$. In optimization theory, this is called the \textit{dual problem} \cite{boyd_2023}:
\begin{equation}
    MD=\max_\lambda LB_L
    \label{dual_problem}
\end{equation}
Combined with the minimization process (\ref{min_L}), the additional maximization procedure (\ref{dual_problem}) shapes the energy landscape to bring the minimum of $L$ towards an optimal feasible point.

Compared to the classical penalty method, adding Lagrange multipliers provides an additional degree of freedom for shaping the energy landscape and closing the gap $G=OPT-LB_L$, as illustrated in Fig.\ref{LR_illustration} with a toy example. Consequently, it is possible to get $G=0$ with a smaller penalty parameter $P<P_C$ which greatly increases the accuracy for hard problems as we show next.

\section{Self-Adaptive Ising Machine}
\subsection{Algorithm}
We now present how to find the optimal Lagrange multipliers $\lambda^*$ in a self-adaptive and iterative manner to get the minimum gap $G$ and find good solutions for (\ref{primal_problem}) during the process. The idea is to alternate the minimization process (\ref{min_L}) and the maximization of the lower bound (\ref{dual_problem}) to bring the system to a feasible region. This mechanism is well-known in optimization and has been used extensively \cite{fisher_2004}.
Since the dual function $LB_L$ is a concave function of Lagrange multipliers $\lambda$ \cite{boyd_2023}, the optimal $\lambda^*$ for the dual problem (\ref{dual_problem}) can be found by ascent in the $\lambda$ space with a subgradient given by
$ \nabla_\lambda LB_L=g(\Bar{x})$ \cite{fisher_2004}. Moreover, the method also converges with a pseudo-minimum of $L$ and is called the surrogate gradient method \cite{xing_zhao_1997}. The latter is useful for IMs since they are heuristic solvers and cannot guarantee to solve exactly (\ref{min_L}) in practice. 

We now combine these techniques and adapt them to Ising machines.
The proposed Algorithm \ref{Alg} works as follows. First, the Lagrange multipliers are set to 0, and the penalty parameter is initialized to some value which can be problem-dependent. For the problems we solve hereafter, we follow the heuristic rule from \cite{Bontekoe_2023,parizy_2021} and set $P=\alpha dN$ where $d$ is the density of the $J$ matrix, $N$ is the number of Ising spins (including slack spins), and $\alpha$ is a constant that can be adjusted for different problems. Then, for a fixed number of iterations $K$, the Lagrange function (\ref{Lagrange_energy}) is minimized by an IM, and the Lagrange multipliers are updated to maximize the lower bound (\ref{dual_problem}) and potentially close the gap. The iteration is similar to an epoch when training a neural network. It shapes the energy landscape such that an optimal solution of the initial problem (\ref{primal_problem}) tends to become a ground state of $L$. In practice, the Ising coefficients $J$ and $h$ (\ref{Ising_hamiltonian}) are consequently updated at each iteration $k$. Meanwhile, feasible solutions $\hat{x_k}$ are stored, and the best one is selected after the for-loop.

\begin{algorithm}[t]
\caption{Self-Adaptive IM for Constrained Optimization} \label{Alg}
\begin{algorithmic}
\Require $f$ and $g$
\Ensure Best feasible solution $(\Bar{x},f(\Bar{x}))$
\State $(\lambda_0,P) \gets (0,\alpha dN)$
\For{K iteration}
\begin{itemize}
    \item Minimize $L_k$: $x_k=\text{argmin}_xL_k$ \Comment{Ising Machine}
    \item Store feasible $\hat{x_k}$ \Comment{CPU}
    \item Update: $\lambda_{k+1}\gets\lambda_k+\eta\, g(x_k)$ \Comment{CPU}
\end{itemize}
\EndFor
\State $\Bar{x}\gets\text{argmin}_k f(\hat{x_k})$
\end{algorithmic}
\end{algorithm}

\subsection{A probabilistic-bit proof-of-concept}
The minimization process of Algorithm \ref{Alg} is compatible with any Ising machine. As a proof of concept, we choose to emulate a probabilistic-bit (p-bit) IM \cite{camsari_2017} in software since p-bit-based architectures are currently very scalable and compatible with various hardware platforms, digital \cite{aadit_2022}, analog \cite{kaiser_2022} or mixed-signal \cite{singh_2024}. A p-computer is composed of stochastic neurons called \textit{p-bit} that are interconnected by weights $J$ and biased by $h$. Each p-bit takes one of the two values $m_i=\pm1$ and receives as input:
\begin{equation}
    I_i=\sum_jJ_{ij}m_j+h_i
    \label{Ii_eq}
\end{equation}
which influences the p-bit state as:
\begin{equation}
    m_i=\text{sign}\big[\tanh{\beta I_i}+\text{rand(-1,1)}\big]
    \label{mi_eq}
\end{equation}
where rand(-1,1) is a random number uniformly distributed between -1 and 1 modeling noise at the p-bit level. $\beta$ is the inverse temperature parameter which sets the slope of the p-bit activation function. Interconnected p-bits satisfying equations (\ref{Ii_eq}) and (\ref{mi_eq}) are known to follow a Boltzmann distribution of state with probability \cite{camsari_2017}:
\begin{equation}
    P\{m\}=\frac{\exp{-\beta L\{m\}}}{\sum_{m}\exp{-\beta L\{m\}}}
    \label{Boltzmann_distrib}
\end{equation}
where $L$ is our Lagrange function (\ref{Lagrange_energy}). In Matlab, we emulate the probabilistic IM by sequentially updating equations (\ref{Ii_eq}) and (\ref{mi_eq}) which corresponds to the Gibbs sampling Monte Carlo method for the probability distribution (\ref{Boltzmann_distrib}) \cite{camsari_2017}. To find good minima of $L$, we anneal the p-bits such as in simulated annealing (SA) \cite{kirkpatrick_1983} with a linear $\beta$-schedule swept from 0 to $\beta_{max}$. For each SA run $k$, we read the last sample of state $\{m\}$ which corresponds to $x_k$ in Algorithm \ref{Alg}. The Lagrange multipliers are then updated from this last sample (and so $J$ and $h$).

\section{Results}
We benchmark SAIM with knapsack problems that are simple to express but generally hard to solve and have many real-life applications such as resource allocation, capital budgeting, satellite management, etc. \cite{wilbaut_2007}. We first study the quadratic knapsack problem (QKP) \cite{billionnet_2004} which is an Ising problem with a linear constraint and has been explored by several works using IMs \cite{parizy_2021,Bontekoe_2023,jimbo_2022,ohno_2024}. Finally, we focus on multidimensional knapsack problems (MKP) that are particular integer linear programs (ILP) with positive coefficients and multiple constraints \cite{chu_1998}. The parameters used in the experiments are listed in Table \ref{parameters}.
\begin{table}[t]
\centering
\caption{Parameters used in QKP and MKP experiments.}
\label{parameters}
\resizebox{0.45\textwidth}{!}{%
\begin{tabular}{@{}cccccc@{}}
\toprule
Experiment & Penalty & MCS/run & Number of runs & $\beta_{max}$ & \multicolumn{1}{l}{$\eta$} \\ \midrule
QKP        & $2dN$   & 1000    & 2000           & 10            & 20                         \\ \midrule
MKP        & $5dN$   & 1000    & 5000           & 50            & 0.05                       \\ \bottomrule
\end{tabular}%
}
\end{table}

\subsection{Quadratic Knapsack Problems}
\begin{figure}[t!]
    \centering
    \includegraphics[width=\linewidth]{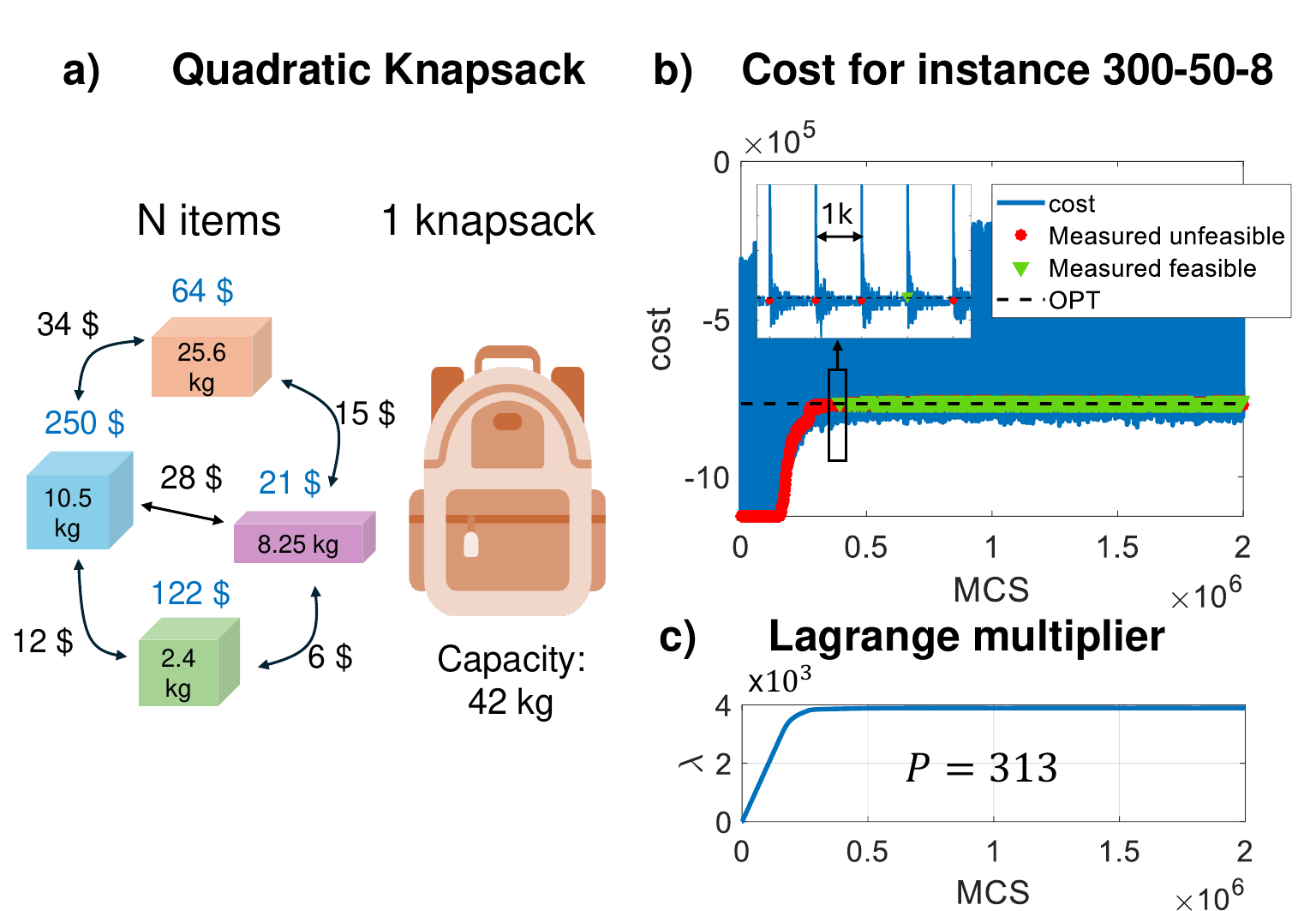}
    \caption{a) The quadratic knapsack problem (QKP) consists of selecting items that maximize individual value (blue \$) and pairwise values (black \$) given a capacity constraint (knapsack). b) Example of simulation result for a 300-variable instance and 50\% of density. c) Corresponding Lagrange multiplier evolution. $\lambda$ remains constant during each SA run of $10^3$ MCS (staircase curve).}
    \label{QKP_illustration}
\end{figure}

Fig.\ref{QKP_illustration}a illustrates QKP which is the generalization of the knapsack problem (NP-hard) where pairs of items also add value to the objective function \cite{billionnet_2004}. It is expressed as:
\begin{align}
   &\min_x -\frac{1}{2}x^T\,W\,x-h^Tx \\ \nonumber
    &x\in \{0;1\}^N \\ \nonumber
    & \text{s.t. } A^T\,x\leq b \nonumber
\end{align}
where $h\in \mathbb{N}^N$ is the value vector for the items, $W$ is a $N$x$N$ positive integer and symmetric matrix representing the additional value when selecting pairs of items, $A\in \mathbb{N}^N$ are the weights of the items and $b\in \mathbb{N}$ the maximum capacity. By using additional slack variables $x_S$, we transform the inequality constraint into equality as $A^Tx+x_S=b$ where $0\leq x_S\leq b$. Using a binary decomposition, $x_S$ is written as $x_S=x_S^0+2x_S^1+...+2^{Q-1}x_S^{Q-1}$ where $x_S^q$ are additional binary variables and $Q=\text{floor}(\log_2(b)+1)$ is the number of additional variables. We include $x_S$ in $x$ of new dimension $N+Q$ and fill $W$ and $h$ with zeros accordingly. The extended vector $A\in\mathbb{N}^{N+Q}$ contains the additional binary coefficients.

We benchmark with instances from \cite{billionnet_2004} that consists of random QKP instances with various $W$-density $d\in\{0.25; 0.5; 0.75;1\}$ and sizes $N\in\{100;200;300\}$. We normalize $W$, $h$ by $\max(|W|,|h|)$ and $A$, $b$ by $\max(|A|,|b|)$ to keep the same $\beta$ schedule for all instances. For every measured sample $x_k$, we check feasibility as $A^Tx_k\leq b$ and if feasible save its cost $c(\hat{x_k})=-\hat{x_k}^TW\hat{x_k}/2-h^T\hat{x_k}$. Since costs are negative, we measure the accuracy (\%) of feasible samples as: 
\begin{equation}
    Accuracy=100\,c(\hat{x_k})/OPT
\end{equation}

\begin{figure}[t!]
    \centering
    \includegraphics[width=0.9\linewidth]{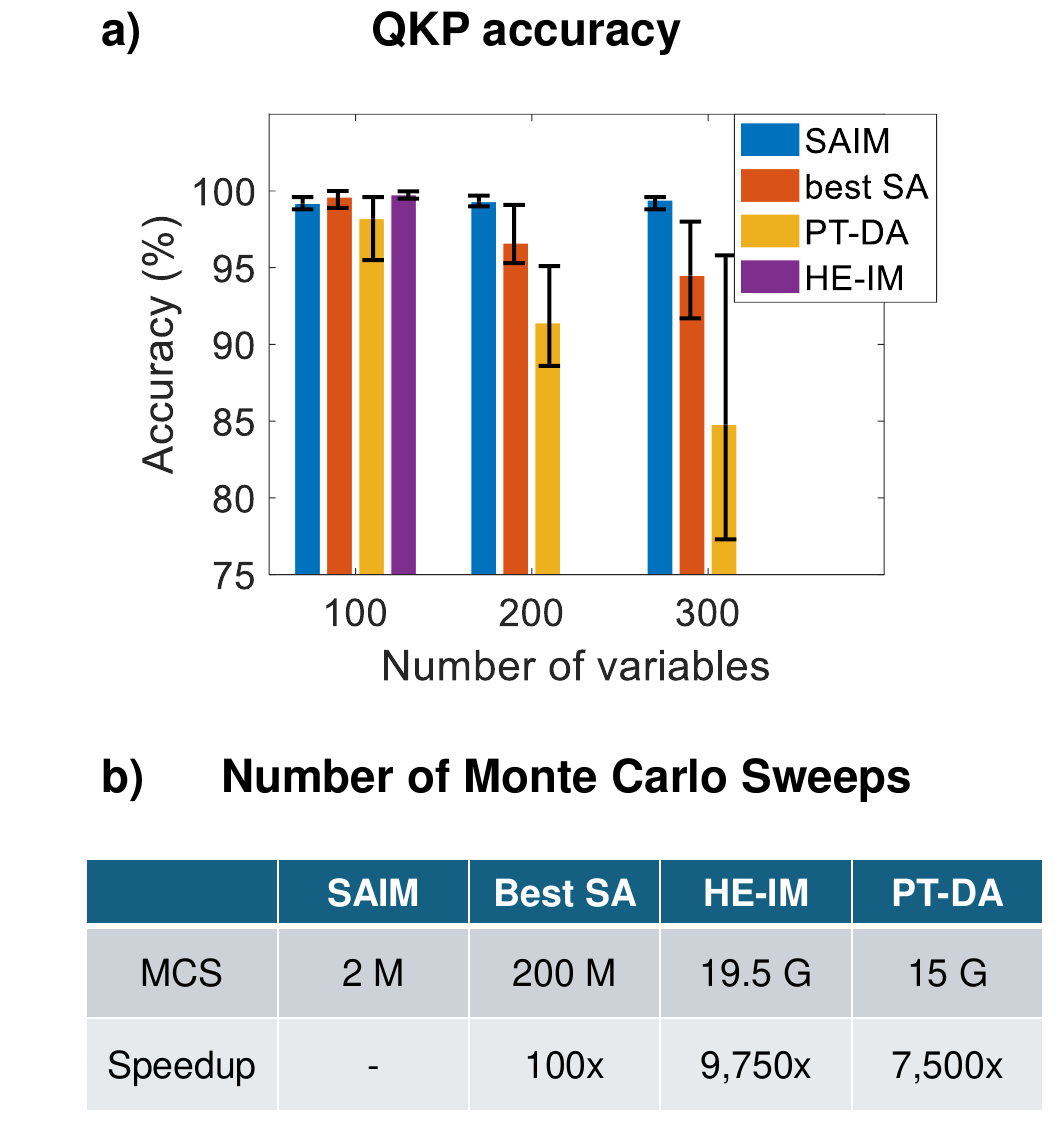}
    \caption{a) QKP results (quartiles). b) Number of reported MCS for each method.}
    \label{QKP_results}
\end{figure}

Fig.\ref{QKP_illustration}b presents an example of cost evolution for the instance 300-50-8. Initially, $\lambda$ is small and the measured samples are all unfeasible (red data points). During this transient time (at the $\lambda$ time scale), the SAIM minimization produces unfeasible samples with a cost $c(x_k)<OPT$, highlighting that the chosen penalty parameter $P=2dN=313$ is too small. However, as shown in Fig.\ref{QKP_results}c, the Lagrange multiplier eventually converges to a steady value $\lambda^*$ for which the IM finds good feasible solutions (green triangles).
\begin{table}[b!]
\centering
\caption{Penalty Method vs. SAIM for QKP.\\ Each instance is named with the prefix $N$-$d$.\\ Parenthesis indicate feasibility.}
\label{SAIM_vs_penalty}
\resizebox{0.5\textwidth}{!}{%
\begin{tabular}{@{}cccccccc@{}}
\toprule
\multicolumn{1}{l}{} & \multicolumn{4}{c}{2000 SA runs of $10^3$ MCS}                    & \multicolumn{3}{c}{10 SA runs of $2\times 10^5$ MCS} \\ \midrule
\multicolumn{1}{l}{} & \multicolumn{2}{c}{SAIM} & \multicolumn{2}{c}{Penalty method} & \multicolumn{3}{c}{Penalty method}         \\ \midrule
Instance    & Best           & Avg                & Best & Avg       & Best & Avg       & Tuned P \\ \midrule
100\_25\_1  & \textbf{100}   & \textbf{99.6 (73)} & 92.0 & 41.0 (87) & 97   & 94.8 (30) & 130dN   \\
100\_25\_2  & \textbf{100}   & \textbf{98.5 (31)} & 83.3 & 36.7 (90) & 82.9 & 71.7 (80) & 60dN    \\
100\_25\_3  & \textbf{100}   & \textbf{98.0 (54)} & 57.5 & 15.5 (96) & 77.0 & 67.1 (50) & 70dN    \\
100\_25\_4  & \textbf{100}   & \textbf{99.2 (66)} & 90.7 & 35.4 (97) & 77.8 & 71.0 (70) & 120dN   \\
100\_25\_5  & \textbf{99.2}  & \textbf{99.2 (37)} & 86.6 & 44.9 (92) & 86.6 & 69.4 (90) & 60dN    \\
100\_25\_6  & \textbf{100}   & \textbf{99.2 (60)} & 94.8 & 36.7 (95) & 98.9 & 91.8 (30) & 250dN   \\
100\_25\_7  & \textbf{100}   & \textbf{99.0 (34)} & 70.9 & 33.9 (94) & 83.4 & 72.2 (40) & 200dN   \\
100\_25\_8  & \textbf{100}   & \textbf{97.3 (24)} & 87.8 & 35.5 (96) & 88.3 & 87.5 (30) & 200dN   \\
100\_25\_9  & \textbf{100}   & \textbf{99.3 (54)} & 96.2 & 37.5 (95) & 83.4 & 79.8 (50) & 300dN   \\
100\_25\_10 & \textbf{99.98} & \textbf{99.3 (54)} & 75.0 & 17.7 (98) & 95.0 & 83.0 (60) & 500dN   \\
100\_50\_1  & \textbf{99.8}  & \textbf{98.9 (49)} & 94.5 & 38.8 (95) & 91.2 & 82.4 (60) & 400dN   \\
100\_50\_2  & \textbf{99.6}  & \textbf{99.5 (68)} & 81.0 & 63.6 (74) & 80.7 & 67.8 (20) & 40dN    \\
100\_50\_3  & \textbf{99.1}  & \textbf{97.4 (28)} & 78.6 & 28.3 (95) & 97.3 & 86.2 (20) & 100dN   \\
100\_50\_4  & \textbf{99.9}  & \textbf{99.8 (82)} & 83.1 & 37.2 (92) & 81.5 & 69.7 (40) & 40dN    \\
100\_50\_5  & \textbf{99.9}  & \textbf{97.5 (12)} & 92.8 & 40.3 (92) & 94.7 & 90.6 (40) & 300dN   \\
100\_50\_6  & \textbf{100}   & \textbf{99.9 (91)} & 68.2 & 21.5 (97) & 75.3 & 66.0 (40) & 200dN   \\
100\_50\_7  & \textbf{99.9}  & \textbf{99.7 (73)} & 88.1 & 24.0 (96) & 99.4 & 97.7 (40) & 220dN   \\
100\_50\_8  & \textbf{99.9}  & \textbf{99.1 (52)} & 89.8 & 25.4 (95) & 99.7 & 93.3 (30) & 220dN   \\
100\_50\_9  & \textbf{100}   & \textbf{99.8 (91)} & 93.1 & 35.2 (94) & 97.4 & 92.1 (40) & 350dN   \\
100\_50\_10 & \textbf{99.4}  & \textbf{99.0 (49)} & 95.0 & 60.2 (94) & 87.6 & 79.0 (50) & 150dN   \\ \midrule
Average     & \textbf{99.8}  & \textbf{99.0 (54)} & 85.0 & 35.5 (93) & 88.8 & 80.7 (47) & 195dN   \\ \bottomrule
\end{tabular}%
}
\end{table}

\begin{table}[t!]
\centering
\caption{QKP results for 200 variables. Optimality is the ratio of optimal solutions over feasible solutions.}
\label{table_200}
\resizebox{0.43\textwidth}{!}{%
\begin{tabular}{@{}ccccc@{}}
\toprule
Instance     & Optimality (\%) & Avg SAIM               & best SA \cite{Bontekoe_2023} & PT-DA \cite{parizy_2021} \\ \midrule
200\_25\_1   & 0               & \textbf{99.3 (34)} & 98.9           & 97.6         \\
200\_25\_2   & 0               & 99.7 (57)          & \textbf{99.9}  & 99.6         \\
200\_25\_4   & 0               & \textbf{99.3 (8)}  & 99.2           & 95.1         \\
200\_25\_5   & 0               & \textbf{99.3 (57)} & 99.1           & 96.7         \\
200\_25\_6   & 0               & \textbf{98.8 (45)} & 96.6           & 92.7         \\
200\_25\_7   & 0               & \textbf{99.3 (53)} & 98.6           & 89.3         \\
200\_25\_8   & 0               & \textbf{99.1 (29)} & 98.3           & 95.6         \\
200\_25\_9   & 1.7             & \textbf{99.3 (57)} & 99.2           & 91.8         \\
200\_25\_10  & 0.1             & 99.0 (53)          & \textbf{99.3}  & 90.5         \\
200\_50\_1   & 100             & \textbf{100 (92)}  & 97.1           & 95.5         \\
200\_50\_2   & 0               & \textbf{99.3 (50)} & 94.9           & 92.6         \\
200\_50\_3   & 0               & \textbf{99.6 (55)} & 95.7           & 81.9         \\
200\_50\_4   & 82              & \textbf{99.8 (80)} & 95.1           & 81.1         \\
200\_50\_5   & 0               & 99.8 (29)          & \textbf{99.9}  & 99.7         \\
200\_50\_6   & 0               & 95.6 (29)          & \textbf{98.7}  & 98.4         \\
200\_50\_7   & 0               & \textbf{98.5 (34)} & 94.5           & 88.6         \\
200\_50\_8   & 6.7             & \textbf{99.3 (58)} & 96.3           & 94.1         \\
200\_50\_9   & 0.3             & \textbf{98.8 (16)} & 97.5           & 88.8         \\
200\_50\_10  & 0               & \textbf{99.1 (55)} & 95.9           & 91.4         \\
200\_75\_1   & 0               & \textbf{99.4 (41)} & 95.5           & 83.5         \\
200\_75\_2   & 0               & \textbf{98.9 (40)} & 91.3           & 84.9         \\
200\_75\_3   & 0               & 98.0 (22)          & \textbf{100}   & 96.0         \\
200\_75\_4   & 23              & \textbf{99.5 (75)} & 86.6           & 81.2         \\
200\_75\_5   & 0               & \textbf{99.5 (62)} & 96.4           & 89.8         \\
200\_75\_6   & 3.8             & \textbf{98.8 (25)} & 95.5           & 78.8         \\
200\_75\_7   & 0               & \textbf{99.8 (74)} & 93.1           & 90.1         \\
200\_75\_8   & 0               & \textbf{99.9 (75)} & 98.1           & 88.9         \\
200\_75\_9   & 0               & \textbf{99.5 (40)} & 95.7           & 85.8         \\
200\_75\_10  & 0               & \textbf{99.1 (48)} & 95.2           & 91.4         \\
200\_100\_1  & 0               & \textbf{99.8 (21)} & \textbf{99.8}  & 92.8         \\
200\_100\_2  & 0               & \textbf{99.6 (37)} & 94.4           & 85.3         \\
200\_100\_3  & 1.1             & 97.4 (70)          & \textbf{100}   & \textbf{100} \\
200\_100\_4  & 0               & \textbf{99.8 (75)} & 97.8           & 91.7         \\
200\_100\_5  & 0               & \textbf{99.8 (59)} & 96.6           & 92.3         \\
200\_100\_6  & 99              & 99.9 (85)          & \textbf{100}   & 88.7         \\
200\_100\_7  & 0               & \textbf{99.7 (68)} & 95.8           & 93.7         \\
200\_100\_8  & 0               & \textbf{99.4 (36)} & 97.5           & 85.8         \\
200\_100\_9  & 0               & \textbf{98.8 (38)} & 95.3           & 93.7         \\
200\_100\_10 & 0               & \textbf{99.5 (41)} & 93.4           & 90.0         \\ \midrule
Average      & 8.1             & \textbf{99.2 (49)} & 96.7           & 90.9         \\ \bottomrule
\end{tabular}%
}
\end{table}
\begin{table}[t!]
\centering
\caption{QKP results for 300 variables}
\label{table_300}
\resizebox{0.45\textwidth}{!}{%
\begin{tabular}{@{}ccccc@{}}
\toprule
Instance    & Optimality (\%) & Avg SAIM               & best SA \cite{Bontekoe_2023} & PT-DA \cite{parizy_2021} \\ \midrule
300\_25\_1  & 0.8             & \textbf{99.8 (73)} & 99.7           & 84.8         \\
300\_25\_2  & 0               & \textbf{98.7 (26)} & 90.3           & 79.2         \\
300\_25\_4  & 0.2             & \textbf{99.3 (46)} & 95.2           & 86.8         \\
300\_25\_5  & 17              & 98.8 (51)          & \textbf{100}   & 94.4         \\
300\_25\_6  & 0.4             & \textbf{99.4 (51)} & 89.3           & 77.1         \\
300\_25\_7  & 0               & \textbf{99.4 (23)} & 96.7           & 95.8         \\
300\_25\_8  & 5.7             & 98.9 (47)          & \textbf{100}   & \textbf{100} \\
300\_25\_9  & 0.5             & \textbf{99.7 (72)} & 90.4           & 87.2         \\
300\_25\_10 & 11              & \textbf{99.7 (59)} & 91.7           & 95.8         \\
300\_50\_1  & 0               & \textbf{99.1 (50)} & 91.2           & 65           \\
300\_50\_2  & 67              & \textbf{99.5 (39)} & 94.4           & 58.3         \\
300\_50\_3  & 0               & \textbf{99.5 (12)} & 95.7           & 82.5         \\
300\_50\_4  & 0               & \textbf{99.5 (57)} & 92.9           & 59.3         \\
300\_50\_5  & 0               & \textbf{99.2 (32)} & 94.3           & 77.8         \\
300\_50\_6  & 0               & \textbf{98.0 (37)} & 94.5           & 77.3         \\
300\_50\_7  & 0               & 98.7 (25)          & \textbf{99.8}  & 97.7         \\
300\_50\_8  & 0               & \textbf{99.6 (36)} & 94.5           & 77.3         \\
300\_50\_9  & 0               & \textbf{99.8 (68)} & 94.9           & 90.7         \\
300\_50\_10 & 0               & \textbf{98.6 (9)}  & 98             & 95.8         \\ \midrule
Average     & 5.4             & \textbf{99.2 (43)} & 94.9           & 83.3         \\ \bottomrule
\end{tabular}%
}
\end{table}

We now compare SAIM and the penalty method with the same total amount of 2M MCS for $N=100$ $d=0.25$ and $d=0.5$ (Table \ref{SAIM_vs_penalty}). For the penalty method, we run 10 SA runs of $2\times 10^5$ MCS each with penalty parameters $P$ tuned in the following manner. An initial small $P=2dN$ was set and coarsely increased until getting a satisfactory feasibility ratio ($\geq20$\%). We note that on average, a large $P$ value implies a feasibility increase, as it has been observed in previous works \cite{jimbo_2022,parizy_2021}. However, we did not find a clear correlation between $P$ values and accuracies. Overall, finding a satisfactory $P$ value is not straightforward and the tuning phase worsens the time-to-solution.

In contrast, the proposed SAIM is less parameter-sensitive as $P$ is set once to $2dN$ for all instances and Lagrange multipliers are automatically updated. Looking at the best-obtained accuracies, we measure an average of 99.8\% for SAIM against 88.8\% for the tuned penalty method. We also tested the penalty method in the same setup as SAIM, i.e. with 2000 SA runs of $10^3$ MCS each. The best accuracy for the penalty method decreased to 85\% on average, highlighting that the high SAIM accuracy does not originate from a large number of SA runs.

Next, we benchmark SAIM with 4 previous works using IMs standalone, i.e. that do not post-process the IM results. The accuracy can be enhanced using heuristics \cite{ohno_2024} but they are problem-dependent and beyond the scope of this paper. All the previous works we have found use the penalty method. We first benchmark with work \cite{Bontekoe_2023} that explores various QUBO encodings and solves QKP with SA. For each instance, we report the best accuracy reported in \cite{Bontekoe_2023} (best SA). We also benchmark with an implementation of the parallel tempering algorithm (using 26 replicas) executed on Fujitsu's Digital Annealer (PT-DA) \cite{parizy_2021}. Finally, SAIM results are compared against the work \cite{jimbo_2022} that uses a hybrid encoding for the slack variables $x_S$ for problems up to 100 variables, and uses SA to find good solutions (HE-IM).

Fig. \ref{QKP_results} summarizes the results obtained with state-of-the-art IMs and Tables \ref{table_200} and \ref{table_300} provide more details for 200 and 300 variables.
The SAIM median accuracy is larger than 99.2\% for all sizes and the solutions are consistently of high quality with interquartile ranges smaller than 0.8\%. In contrast for 300 variables, the median accuracies for the best SA algorithm \cite{Bontekoe_2023} and PT-DA \cite{parizy_2021} are 94.5\% and 84.8\%, respectively. 

SAIM requires much fewer samples, as shown in Fig.\ref{QKP_results}b. The best SA method from \cite{Bontekoe_2023} has reported 200,000 trials of 1000 MCS. The HE-IM \cite{jimbo_2022} used 26 trials of 750 M MCS. PT-DA \cite{parizy_2021} reported 20 trials of 750 M MCS. Overall, SAIM requires 100x fewer samples than the best SA runs and 7,500x fewer than PT-DA.

\subsection{Multidimensional Knapsack Problems}
\begin{figure}[t!]
    \centering
    \includegraphics[width=\linewidth]{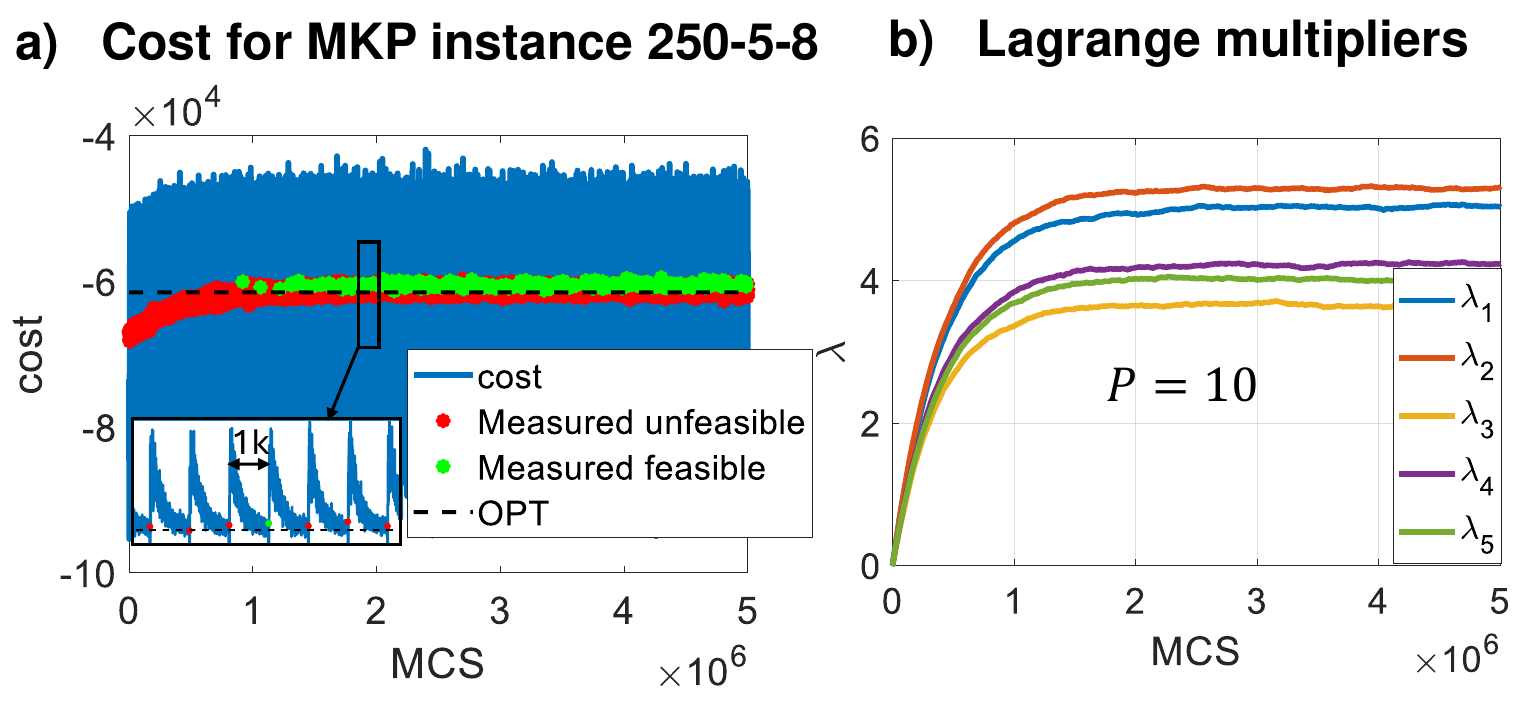}
    \caption{ a) Example of MKP simulation for a 250-variable instance and 5 knapsacks. b) Corresponding Lagrange multipliers dynamics for a fixed $P=10$. $\lambda$ remain constant during each SA run of $10^3$ MCS (staircase curve).}
    \label{MKP_example}
\end{figure}
To test the applicability of SAIM with multiple constraints, we focus on the NP-hard MKP which consists of selecting items that are subjected to several capacity constraints (knapsacks) and expressed as \cite{wilbaut_2007}:
\begin{align}
   &\min_x -h^Tx \\ \nonumber
    &x\in \{0;1\}^N \\ \nonumber
    & \text{s.t. } A\,x\leq B \nonumber
\end{align}
where $h\in \mathbb{N}^N$ is the value vector for the items, $A$ is a positive integer $M$x$N$ matrix representing the item weights, and $B\in \mathbb{N}^M$ the vector of maximum capacities. As for QKP, we transform the inequality constraints into equality constraints and normalize $h$, $A$, and $B$ similarly. Since there are no quadratic interactions ($J$ matrix), we approximate the problem density $d$ as $N/(0.5N(N+1))=2/(N+1)$ as if the external fields $h$ were pairwise connections from an additional fixed spin reference to the $N$ initial spins. Compared to QKP, we set $P=5dN$ rather than $2dN$ to compensate the lack of quadratic interaction in the initial cost function.

We benchmark SAIM with a state-of-the-art genetic algorithm for MKP (GA) \cite{chu_1998} using instances with $N\in\{100;250\}$ and $M\in\{5;10\}$.
Fig.\ref{MKP_example} shows an example of SAIM simulation for 250 items and 5 knapsacks. Initially, the constraints are unsatisfied with $A^Tx_k>B$ (the total weights exceed knapsack capacities), and the five corresponding Lagrange multipliers $\lambda$ are updated after each iteration (they increase since $A^Tx_k-B\geq 0$). Then, after approximately 1000 updates, $\lambda$ starts to stabilize and SAIM finds near-optimal solutions.

Table \ref{MKP_results} shows the detailed results. Optimal solutions are obtained with a branch and bound (B\&B) algorithm (\textit{intlinprog} Matlab function) and its execution time on a standard laptop is reported to estimate the instance difficulty. For three various classes of problems, the average best SAIM accuracy is 99.7\%. Although the authors in \cite{chu_1998} reported a mean lower bound of 99.1\% for the GA accuracy, this suggests SAIM solutions are comparable with GA. Obtaining a similar accuracy is encouraging since SAIM does not \textit{a priori} harness the problem structure whereas the proposed GA \cite{chu_1998} is tailored for MKP.

However, the proportion of feasible samples for MKP (5.1\%) is severely reduced compared to the previous QKP study (around 50\%). We believe this is mainly because multiple constraints are harder to satisfy simultaneously. To increase feasibility, one could increase the initial penalties set by $P$. Another approach proposed in \cite{Bontekoe_2023} would be to reduce the knapsack capacities $B$ artificially as $B'<B$ so that the measured samples are more likely to satisfy the constraints.

\begin{table}[t]
\centering
\caption{MKP results. Instances are named with the prefix $N$-$M$.}
\label{MKP_results}
\resizebox{0.45\textwidth}{!}{%
\begin{tabular}{@{}cccccc@{}}
\toprule
\multicolumn{1}{l}{} & \multicolumn{1}{l}{} & \multicolumn{3}{c}{SAIM}              & GA \cite{chu_1998}                            \\ \midrule
Instance             & B\&B time (s)        & Optimality (\%) & Best  & Avg         & Avg                            \\ \midrule
100\_5\_1            & 34                   & 0.5             & 100   & 98.8 (7.5)  & \multirow{10}{*}{$\geq$99.1}   \\
100\_5\_2            & 8.5                  & 16.4            & 100   & 99.1 (7)    &                                \\
100\_5\_3            & 13                   & 0               & 99.9  & 99.4 (6)    &                                \\
100\_5\_4            & 42                   & 0               & 99.8  & 98.3 (7.9)  &                                \\
100\_5\_5            & 25                   & 0               & 99.9  & 98.8 (7.7)  &                                \\
100\_5\_6            & 7                    & 1.6             & 100   & 99.0 (7.7)  &                                \\
100\_5\_7            & 5                    & 2.9             & 100   & 98.8 (8.3)  &                                \\
100\_5\_8            & 20                   & 1.6             & 100   & 98.9 (11.6) &                                \\
100\_5\_9            & 9                    & 0.2             & 100   & 98.9 (8.5)  &                                \\
100\_5\_10           & 20                   & 1.9             & 100   & 98.8 (7.2)  &                                \\ \midrule
100\_10\_1           & 425                  & 0               & 99.97 & 98.4 (2.5)  & \multirow{10}{*}{$\geq$98.4}   \\
100\_10\_2           & 364                  & 0               & 99.6  & 98.5 (2.1)  &                                \\
100\_10\_3           & 189                  & 0               & 99.7  & 98.1 (2.1)  &                                \\
100\_10\_4           & 468                  & 0               & 99.9  & 97.6 (2.2)  &                                \\
100\_10\_5           & 172                  & 0               & 99.3  & 97.7 (2.3)  &                                \\
100\_10\_6           & 899                  & 0               & 99.4  & 97.6 (2.2)  &                                \\
100\_10\_7           & 110                  & 1.2             & 100   & 98.6 (7.6)  &                                \\
100\_10\_8           & 84                   & 0               & 99.6  & 97.8 (1.6)  &                                \\
100\_10\_9           & 70                   & 0               & 99.6  & 98.2 (2.1)  &                                \\
100\_10\_10          & 76                   & 0               & 99.98 & 98.0 (1.8)  &                                \\ \midrule
250\_5\_1            & 327                  & 0               & 99.5  & 98.5 (5.1)  & \multirow{10}{*}{$\geq$99.8}   \\
250\_5\_2            & 617                  & 0               & 99.6  & 98.5 (5.2)  &                                \\
250\_5\_3            & 29                   & 0               & 99.4  & 98.4 (4.7)  &                                \\
250\_5\_4            & 1650                 & 0               & 99.3  & 98.2 (4.4)  &                                \\
250\_5\_5            & 777                  & 0               & 99.5  & 98.4 (4.8)  &                                \\
250\_5\_6            & 1272                 & 0               & 99.5  & 98.4 (4.7)  &                                \\
250\_5\_7            & 348                  & 0               & 99.5  & 98.4 (5)    &                                \\
250\_5\_8            & 1580                 & 0               & 99.7  & 98.3 (5)    &                                \\
250\_5\_9            & 163                  & 0               & 99.7  & 98.6 (4.4)  &                                \\
250\_5\_10           & 49                   & 0               & 99.4  & 98.2 (4.7)  &                                \\ \midrule
Average              & 328                  & 0.9             & \textbf{99.7}  & 98.4 (5.1)  & \multicolumn{1}{l}{$\geq$99.1} \\ \bottomrule
\end{tabular}%
}
\end{table}

\section{Conclusion}
This paper introduces a self-adaptive Ising machine (SAIM) for constrained optimization that iteratively shapes its energy landscape using a Lagrange relaxation of constraints to avoid tuning energy penalties. Emulated with a probabilistic-bit IM in software, we benchmark SAIM for hard quadratic knapsack problems (QKP) and multidimensional knapsack problems (MKP) with multiple constraints. For QKP, SAIM finds better solutions than state-of-the-art IMs such as a parallel tempering algorithm executed on Fujitsu's Digital Annealer and produces 7,500x fewer samples. Compatible with any programmable IM, SAIM has the potential to significantly speed up IMs for constrained optimization.

\section*{Acknowledgment}
We acknowledge support from ONR-MURI grant N000142312708, OptNet: Optimization with p-Bit Networks, and we thank Kerem Yunus Camsari for the enriching discussions.

\section*{Data Availability}
Benchmark data and Matlab codes are available at the following GitHub repository: \url{https://github.com/corentindelacour/self-adaptive-IM}
\bibliography{mybib.bib}{}
\bibliographystyle{IEEEtran}

\end{document}